\documentclass[12pt,preprint]{aastex}
\usepackage{graphicx}

\shorttitle{Dark Matter Clumps and Disk Heating}
\shortauthors{E. Ardi, T. Tsuchiya \& A. Burkert}

\begin{document}

\title{Constraints of the Clumpyness of Dark Matter Halos Through
  Heating of the Disk Galaxies}

\author{Eliani Ardi\altaffilmark{1}, Toshio Tsuchiya}
\affil{Astronomisches Rechen-Institut, M\"{o}nchhofstr 12-14, D69120
  Heidelberg, Germany}
\email{eliani@ari.uni-heidelberg.de, tsuchiya@ari.uni-heidelberg.de}
\and
\author{Andreas Burkert}
\affil{Max-Planck Institut f\"ur Astronomie, Heidelberg, Germany}

\altaffiltext{1}{Permanent Address : Department of Astronomy, Institute of
 Technology Bandung, Indonesia}

\email{burkert@mpia-hd.mpg.de}

\begin{abstract}

 Motivated by the presence of numerous dark matter clumps in the Milky
 Way's halo as expected from the cold dark matter cosmological model,
 we conduct numerical simulations to examine the heating of the disk. We
 construct an initial galaxy model in equilibrium, with a stable thin disk.
 The disk interacts with dark matter clumps for
 about 5 Gyr. Three physical effects are examined : first the mass
 spectrum of the dark matter clumps, second the initial thickness of
 the galactic disk, and third the spatial distribution of
 the clumps. We find that the massive end of the mass spectrum
 determines the amount of disk heating. Thicker disks suffer less
 heating. There is a certain thickness at which the heating owing to the
 interaction with the clumps becomes saturates. We also find that the
 heating produced by the model which mimics the distribution found in
 Standard CDM cosmology is significant and too high to explain the
 observational constraints. On the other hand, our model that corresponds to
 the clump distribution in a $\Lambda$CDM cosmology produces no significant
 heating. This result suggests that the $\Lambda$CDM cosmology is
 preferable with respect to the Standard CDM cosmology in explaining the
 thickness of the Milky Way. 

\end{abstract}


\section{Introduction}

Hierarchical clustering governed by cold dark matter (CDM) is widely
believed as a cosmological scenario which is responsible for the growth
of the structures in the universe. According to the hierarchical scenario,
small dark matter halos should collapse earlier, but later fall into
larger structures. The process of smaller halos being assembled into a
larger halo does not always destroy the smaller ones, thus hierarchical
structures are seen in many objects, such as clusters of galaxies.

Recent high-resolution simulations have successfully shown that hundreds
of galaxy-size DM halos survive in clusters of galaxies \citep{ghig1998,
coli1999, klyp1999a}. A remarkable outcome of the
high-resolution cosmological simulation in Standard CDM model by
\citet{moor1999} even shows that survival of substructures or satellites
occurs not only on cluster scales, but also on galactic scales. They
show that a galaxy of a similar size as the Milky Way should contain
about 500 satellites, which is, however, much more than the number of
the observationally identified satellites.  That many
satellites should survive, was confirmed also by \citet{klyp1999b} and
\citet{ghig2000}. Klypin et al. note that the results of the Standard
CDM simulation are close to those of a $\Lambda$CDM simulation with the
same circular velocity function of substructures. This indicates that
the prediction of the presence of many satellites is a general outcome
of the hierarchical scenario and not particularly dependent on the
cosmological models.

Compared with the observational results, these cosmological models yield
a large number of the satellites, approximately a factor of~50 more
than the number of satellites observed in the vicinity of the Milky Way.

\citet{klyp1999b} suggest that the problem of the missing satellites
could be resolved (i) by identification of some satellites with the
high-velocity clouds observed in the Local Group \citep{blit1999} or
(ii) by considering dark satellites that failed to gather enough gas to
form stars, because of expulsion of gas the supernova-driven winds or
because of gas heating by the intergalactic ionizing background. The latter
possibility implies that the halos of galaxies may contain substantial
substructures in form of numerous invisible DM clumps.

A statistic of strong gravitational lensing is an approach to identify
the dark clumps in the Milky Way. \citet{chib2002} indeed finds evidence for
the existence of the numerous satellites in the Milky Way.

If a great amount of the dark satellites exist within the Milky Way's
halo, their interaction with the disk might cause disk heating.  In
their Standard CDM model, \citet{moor1999} found that a large fraction
of the substructures have very eccentric orbits, so that they could
cause resonant heating of the disk, and even heating by impulsive
shocking owing to their penetration through the disk.

On the other hand, it is known that the Milky Way has a quite thin disk,
whose scale height is about 200 pc. From the ``thinness''
\citet{toth1992} have made an energetical analysis of the disk heating owing
to accretion of matter, and derived the constrain that the Milky Way should
have accumulated no more than 10 \% of the disk mass within the past
5~Gyr. This estimation of the disk heating might, however, be too large
because the actual interaction between the disk and satellites is more
complicated. For example, a single satellite could dissolve before
reaching the disk \citep{huan1997} or the energy injected into
the disk could excitate coherent warping motions rather than heat the
disk \citep{vela1999}. For the case of the interaction of
the disk with many substructures in the halo, additional detailed
numerical investigations are necessary. \citet{font2001} have studied
the case of the $\Lambda$CDM cosmological model, and found that the
substructures are not efficient perturbers for heating of the disk,
since the masses of the clumps are lower than those of the clumps
predicted in the Standard CDM model, and also because the clumps are
located far away from the disk and seldom get near the disk.

The disk kinematics and dynamics is a good probe not only for examining
the cosmological models, but much more generally for clarifying the halo
substructure that is difficult to derive from direct
observations. Therefore in this study we aim to investigate disk heating
by dark matter (DM) clumps for a wide range of parameters. 

Numerical experiments on the disk dynamics are not an easy task,
especially when studying the vertical structure, because of the wide
range of the dynamical scales among the different components in
galaxies. The smallest scale is disk, with a scale height about 200~pc,
while the largest scale is the dark halo extent of $\gtrsim 100$
kpc. Many numerical studies consider disks of 700~pc ($0.2\times$ the
disk scale length) in thickness \citep{vela1999, font2001}. The question
is whether the heating rate obtained for such thick disks is the same as
for thin disks like the real Galactic disk.

In this study we construct our initial galactic models following 
\citet{kuij1995}, which is nearly in exact equilibrium and which
allows us to set up disks as thin as the real Galactic disk. Several
additional observational constraints are taking into account to build
the galactic models. We study three physical effects that could affect
disk heating; first the mass spectrum of the DM clumps, second the
initial thickness of the galactic disks, and the third the spatial
distribution of the clumps.

This paper is organized as follows. In section~2 we present the galaxy
model which provides a very stable thin disk comparable to the real Milky Way
disk. A model of a clumpy dark matter halo is presented in section~3. Numerical
simulations of the interaction between disk and the clumps
are specified in section~4. Section~5 presents the results of our
numerical simulations on the disk heating by examining the effects on
the mass spectrum of the clumps, the initial disk thickness, and the
spatial distribution of the clumps. We summarize and discuss our results
in section~6.

\section{Galaxy Model}

We use the self-consistent disk-bulge-halo galaxy model given by Kuijken
and Dubinski (1995, hereafter KD). This model provides nearly exact
solutions of the collisionless Boltzmann and the Poisson equations, so
that one can control subtle differences in the initial conditions. This
is important especially to set up very thin equilibrium disks, which is
the most difficult part. Using this model we could successfully realize
an equilibrium disk with realistic thickness of only 200~pc.

Here we summarize the model properties briefly. The configurations of
the disk, bulge and halo are determined by their distribution functions.
The bulge distribution function depends only on the relative energy $E$,
while that of the halo depends also on the angular momentum about the
symmetric axis, $L_z$. 

The halo distribution function takes the form
\begin{equation}
  \label{eq:haloDF}
  f_{\mathrm{halo}}(E,L_z^2) = \left\{
    \begin{array}{ll}
      \left[ (AL_z^2+B)\exp (-E/\sigma_0^2) + C\right]
      \left[ \exp (-E/\sigma_0^2) -1\right]  \quad&
      \mathrm{if } E<0, \\
      0 & \mathrm{otherwise}, \\
    \end{array}
    \right.
\end{equation}

The relative energy $E$ is defined so that $E=0$ at the
edge of the distribution, where $\rho=0$. The halo distribution function
has five free parameters: the potential at the
center $\Psi_0$, which appears implicitly in the definition of the
relative energy, the velocity scale $\sigma_0$, and three factors $A$,
$B$, and $C$. The factors $A$ and $B$ control the system flattening
($q$) and the core radius ($R_c$), respectively, and all the three factors
are scaled by the density scale ($\rho_1$).

The bulge distribution function takes the same form of a King model
\citep{binn1987}which is given by
\begin{equation}
  f_{\mathrm{bulge}}(E)=\left\{ 
    \begin{array}{ll}
      \rho_\mathrm{b}(2\pi\sigma_\mathrm{b}^2)^{-3/2}
      \exp[(\Psi_0-\Psi_\mathrm{c})/\sigma_\mathrm{b}^2] &
      \{\exp[-(E-\Psi_\mathrm{c})/\sigma_\mathrm{b}^2]-1\} \\
      & \mbox{if } E<\Psi_\mathrm{c}, \\
      0 & \mbox{otherwise}.
    \end{array} \right.
\end{equation}
This distribution depends on three parameters: the cutoff potential
$\Psi_\mathrm{c}$, the center density $\rho_\mathrm{b}$, and the
velocity dispersion $\sigma_\mathrm{b}$.

 For halo and bulge distributions, the densities are given by
analytic functions of $R$ and $\Psi$.

The disk distribution function must depend on three integrals of motion
so that it can sustain a triaxial velocity ellipsoid within the disk.
Since the third integral is not analytic, KD employ an
approximate form of the distribution function which depends on $E$,
$L_z$ and the vertical energy $E_z\equiv
\Psi(R,z)-\Psi(R,0)+\frac{1}{2}v_z^2$.  $E_z$ is only an approximate
integral, but well conserved for stars in nearly circular orbits in a
warm disk. This quantity is the only approximation in KD's model. The
distribution function takes the form
\begin{equation}
  \label{eq:diskDF}
  f_{\mathrm{disk}}(E_\mathrm{p},L_z,E_z) =
  \frac{\Omega(R_\mathrm{c})}{(2\pi^3)^{1/2}\kappa(R_\mathrm{c})}
  \frac{\tilde{\rho_\mathrm{d}}(R_\mathrm{c})}
  {\tilde{\sigma_R}^2(R_\mathrm{c})\tilde{\sigma}_z(R_\mathrm{c})}
  \exp\left[ -\frac{E_p-E_\mathrm{c}(R_\mathrm{c})}
  {\tilde{\sigma}_R^2(R_\mathrm{c})} - 
  \frac{E_z}{\tilde{\sigma}_z^2(R_\mathrm{c})} \right]
\end{equation}
where $E_\mathrm{p}\equiv E-E_z$, $R_\mathrm{c}$ and $E_\mathrm{c}$ are
the radius and energy of a circular orbit with angular momentum
$L_z$. $\Omega$ and $\kappa$ are the circular and epicyclic frequencies
at radius $R_\mathrm{c}$, respectively. The `tilde' functions
$\tilde{\rho_\mathrm{d}}$, $\tilde{\sigma}_R$, and $\tilde{\sigma}_z$
are free functions. Therefore the density and the radial velocity
dispersion are conveniently selected as
\begin{equation}
  \label{eq:disk_density}
  \rho_\mathrm{disk}(R,z) = \frac{M_\mathrm{d}}{8\pi
    R_\mathrm{d}^2z_\mathrm{d}}\mathrm{e}^{-R/R_\mathrm{d}}
    \mathrm{erfc} \left(
    \frac{r-R_\mathrm{out}}{\sqrt{2}\delta R_\mathrm{out}} \right)
  \exp\left[ -4.6187\frac{\Psi_z(R,z)}{\Psi_z(R,3z_\mathrm{d})}
    \right],
\end{equation}
and
\begin{equation}
  \label{eq:disk_rad_velocity_dispersion}
  \tilde{\sigma}_R^2 = \sigma_R^2(0)\exp(-R/R_\mathrm{d}).
\end{equation}
$\tilde{\rho_\mathrm{d}}$ and $\tilde{\sigma}_z$ are iteratively
adjusted so that the densities on the mid-plane and at height
$z=3z_\mathrm{d}$ will agree with those given by eq.~(\ref{eq:disk_density}).
The distribution of the disk has six free parameters: the disk mass
$M_\mathrm{d}$, the radial scale length $R_\mathrm{d}$, the vertical
scale height $z_\mathrm{d}$, the disk truncation radius
$R_\mathrm{out}$, the truncation width $\delta R_\mathrm{out}$, and the central
velocity dispersion of the disk $\sigma_R(0)$.

The choice of the above parameters is made to satisfy observational
properties of the Milky Way (see also \citet{tsuc2002}), that is the
solar radius $R_0$ = 8 kpc, the circular velocity of the disk at the
solar radius $V_\mathrm{c}=220$ km/s, the total surface density within
1.1 kpc of the disk plane $\Sigma_{1.1}(R_0) = 71 \pm 6 M_\odot
\mathrm{pc}^{-2}$ \citep{kuij1991}, the contribution of the
disk material to $\Sigma_{1.1}$, $\Sigma_\mathrm{d}(R_0) = 48 \pm 9
M_\odot \mathrm{pc}^{-2}$ \citep{kuij1991}, the total Galaxy
mass within 50 kpc $M_\mathrm{tot}(<50\mathrm{kpc}) = 5.4^{+0.2}_{-3.6}
\times 10^{11} M_\odot$ \citep{wilk1999}, and the total Galaxy
mass within 170 kpc $M_\mathrm{tot}(<170\mathrm{kpc}) =
1.9^{+3.6}_{-1.7} \times 10^{12} M_\odot$ \citep{wilk1999}. 

\begin{figure}
  \plotone{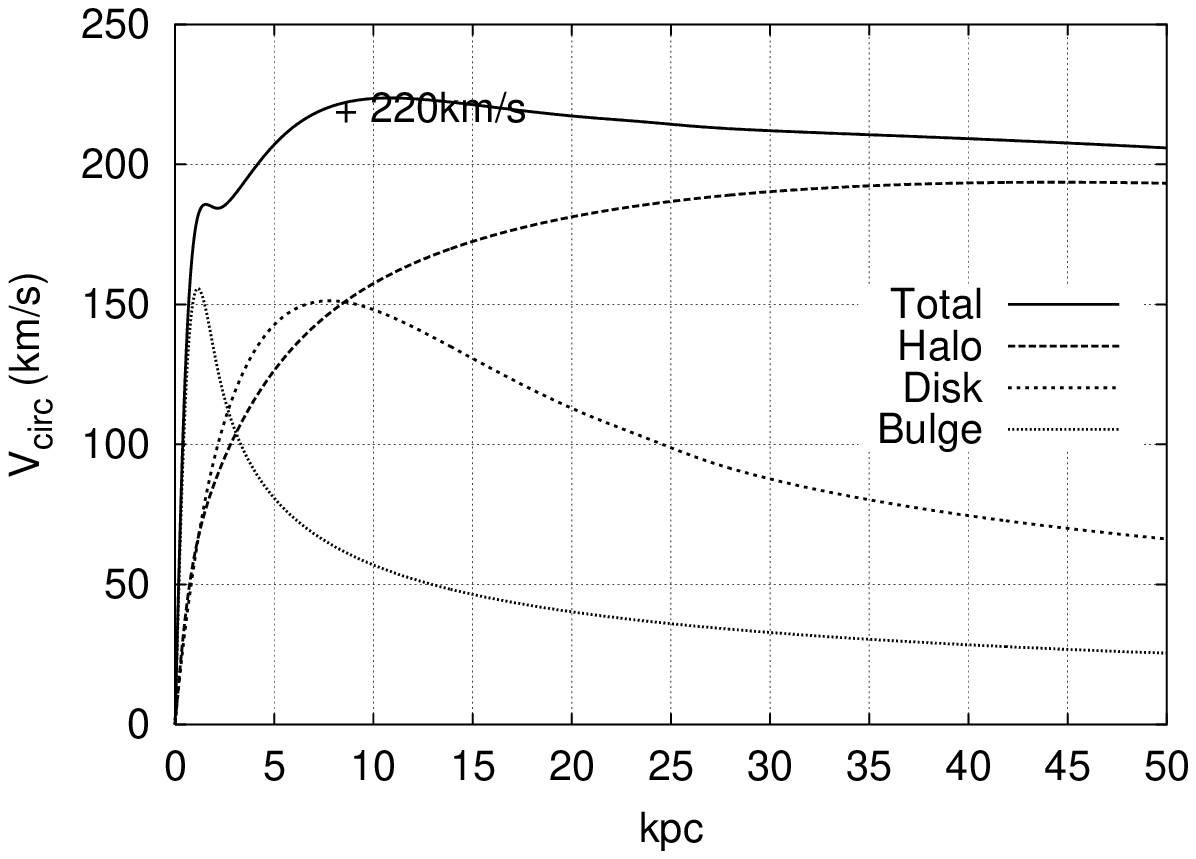}
  \caption{Circular velocity profiles for the Galaxy model with the
    standard halo. The cross sign (+) shows the observational circular
    velocity at the solar radius.\label{fig:modelF_circ}}
\end{figure}

We assume that the disk mass is $M_\mathrm{d}=5\times 10^{10}M_\odot$,
the disk scale length $R_\mathrm{d}=3.5$~kpc, the disk scale height
$z_\mathrm{d}=245$~pc, and the disk truncation radius and the truncation
width are $R_\mathrm{out}=7R_\mathrm{d}$ and $\delta R=R_\mathrm{d}$,
respectively. The bulge mass is about 15 \% of that of the disk,
while the tidal radius is 2.38~kpc. The contributions of the bulge and
disk to the rotation curve is shown in Fig.~\ref{fig:modelF_circ}. We
also examined thicker disks with $z_\mathrm{d}=525$~pc and 1050~pc, as
described in the next section.

\begin{figure}
  \plotone{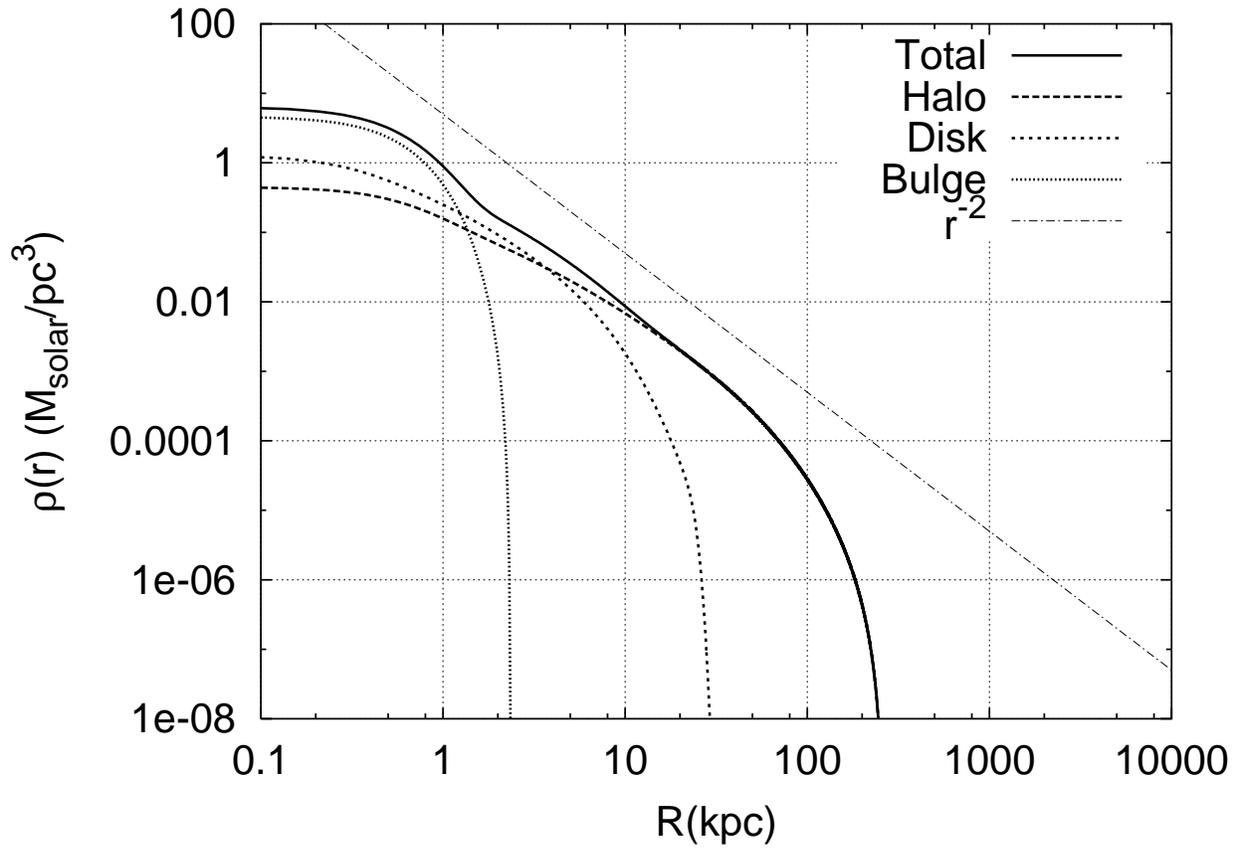}
  \caption{Volume density profiles of the components in the Galaxy model
    with the standard halo.\label{fig:modelF_dens}}
\end{figure}

\begin{figure}
  \plotone{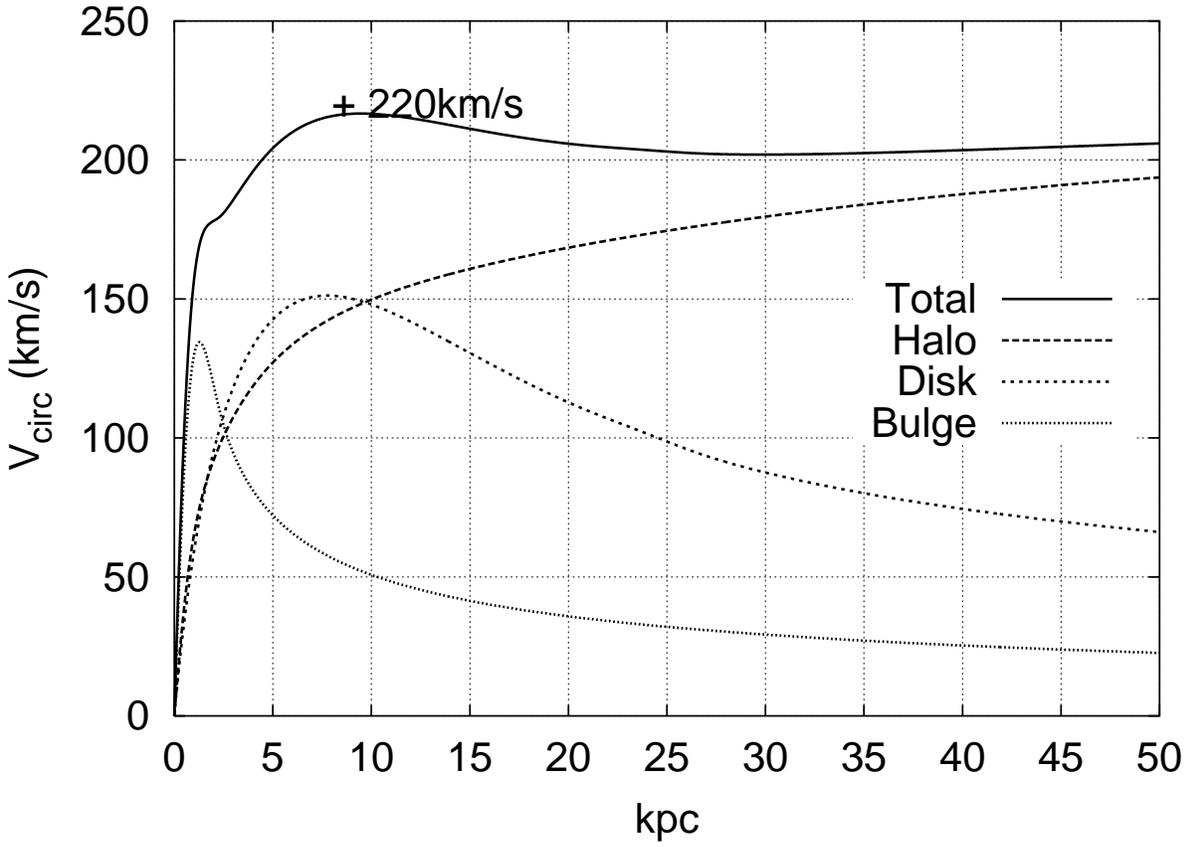}
  \caption{Circular velocity profiles for the extended halo model. The cross
    sign (+) shows the observational circular velocity at the solar
    radius.\label{fig:modelF+_circ}}
\end{figure}

\begin{figure}
  \plotone{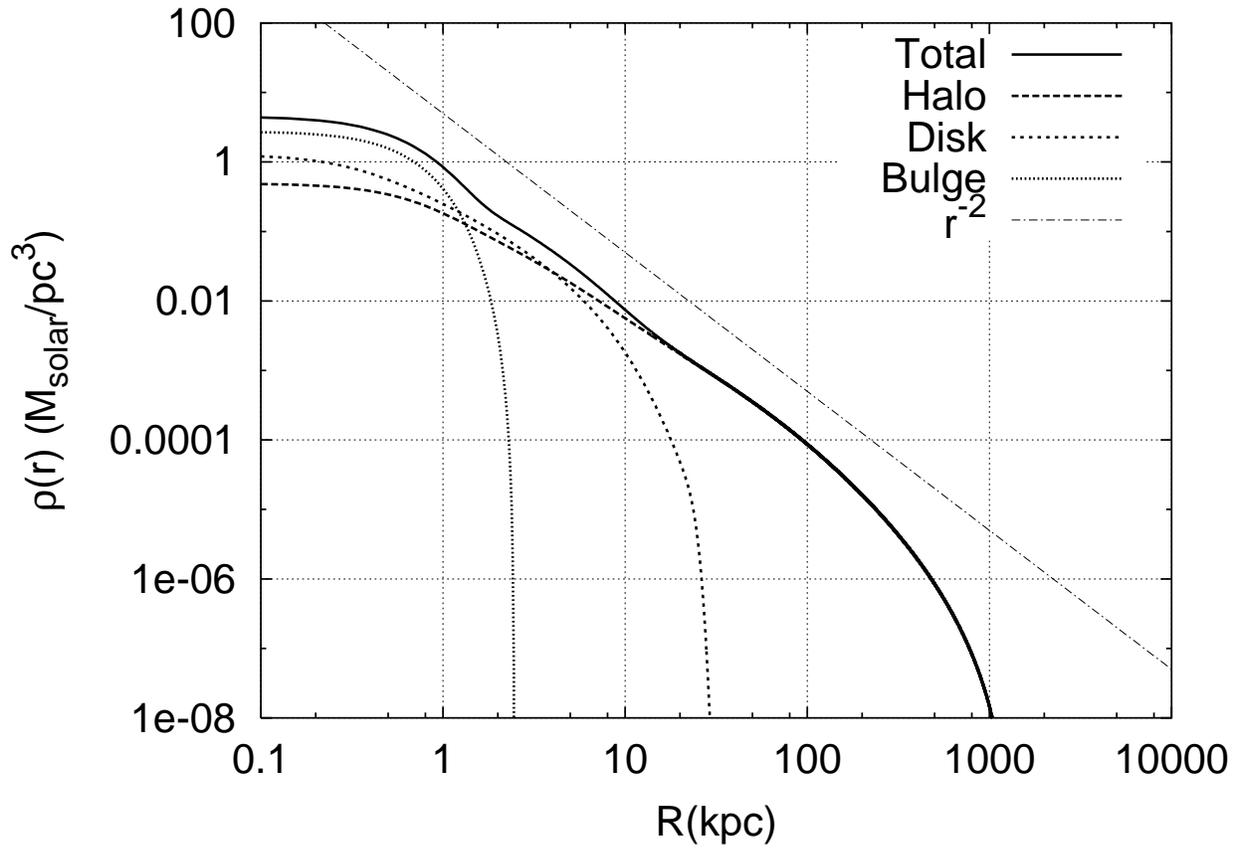}
  \caption{Volume density profiles of the components in the extended
  halo model.\label{fig:modelF+_dens}}
\end{figure}

For the halo, we assume $A$=0 in eq.~\ref{eq:haloDF}, which means that
the halo is nearly spherical, except for the flatness caused by the
presence of the disk potential. As for the extent of the halo, we
consider two different models. One has the virial radius at 262~kpc,
with a dark matter mass of $8.59 \times 10^{11}M_{\odot}$. We refer to
it as the \emph{standard halo model}. The other has the virial radius at
1.35~Mpc, with a mass of $4.1\times 10^{12}M_{\odot}$, which is called
the \emph{extended halo model}. The difference between the two models
lies only in the outer part of the halos. The inner density profiles ($R
\lesssim 50$~kpc) are about the same (Fig.~\ref{fig:modelF_dens} and
Fig.~\ref{fig:modelF+_dens}), thus the rotation curves of the bulge,
disk, and the halo are nearly equivalent between two models
(Fig.~\ref{fig:modelF_circ} and Fig.~\ref{fig:modelF+_circ}).  The
extended halo model is introduced only to produce different clump
distributions as described below.

The halo and the bulge are treated as an external force, while the
distribution of the disk is realized by self-consistent particles.

Without the substructure clumps in the halo, this model is fairly
stable. The net increase in the disk thickness is only due to 2-body
relaxation among the disk particles, which is only $\Delta z_d\sim
150$~pc after 4.76 Gyr.

\section{Dark Matter Clumps Model}

The numerical cosmological simulations \citep{moor1999} have demonstrated the
presence of the dark matter clumps within dark halos. We
distribute the clumps in a similar way as shown in the numerical results.

The dark matter clumps are represented by rigid bodies with a 
Navarro, Frenk \& White's density profile \citep{nava1997}. It
provides an accurate fit to the density profiles of CDM halos:
\begin{equation}
 \label{eq: NFW_dens1}
 \rho(r)= \frac{\rho_{\rm crit}{\delta_{\rm c}}}{(r/r_\mathrm
 {s}){(1+r/r_{\mathrm {s}})}^2}. 
\end{equation}
Here $\rho_{\rm crit}= 3 H^2/8 \pi G$ is the critical density for
closure of the universe, and $\delta_{\rm c}$ and $r_\mathrm{s}$ are the
characteristic density and the scale radius of the clump. In practice,
it is more convenient to use mass and concentration to characterize each
clump instead of $\delta_\mathrm{c}$ and $r_\mathrm{s}$.

The mass of a dark matter clump, $M$, is related to its virial
radius $r_{200}$, which is defined as the radius within which the
average density equals 200 times the critical density at the present
$\rho_\mathrm{crit,0}$, by the relation
\begin{equation}
  M = \frac{4\pi r_{200}^3}{3} 200 \rho_\mathrm{crit,0}.
\end{equation}
Then the characteristic density $\delta_{\rm c}$ is expressed in terms
of the concentration $c=r_{200}/r_s$ as
\begin{equation}
\delta_{\rm c}=\frac{200}{3} \frac{c^3}{\ln(1+c)-c/(1+c)}.
\end{equation}

Each clump has two characteristic parameters, but we fix the
concentration to a typical value $c=20$. Since the interaction between
the clumps and the disk is dominated by distant encounters, the adopted
value for the concentration does not affect the heating process so much.

The simulations of \citet{klyp1999b} and \citet{ghig2000} show
that the total mass of the clumps is about one tenth of the halo mass.
We therefore give the clumps 10 \% of the halo
mass of the standard model, which is $8.59 \times 10^{10}M_{\odot}$ in
total. The distribution of positions and velocities of the clumps is
assumed 
to be the same as the halo distribution function (eq.\ref{eq:haloDF}),
so that the clump distribution does not change in time. This is
the most reasonable assumption which is important in order to maintain
an equilibrium in the whole system. Otherwise
the change in the mean potential of the halo might cause undesired
change in the disk thickness. The background halo component, which is
treated as an fixed external field, is reduced by 10 \%.

\begin{figure}
      \plotone{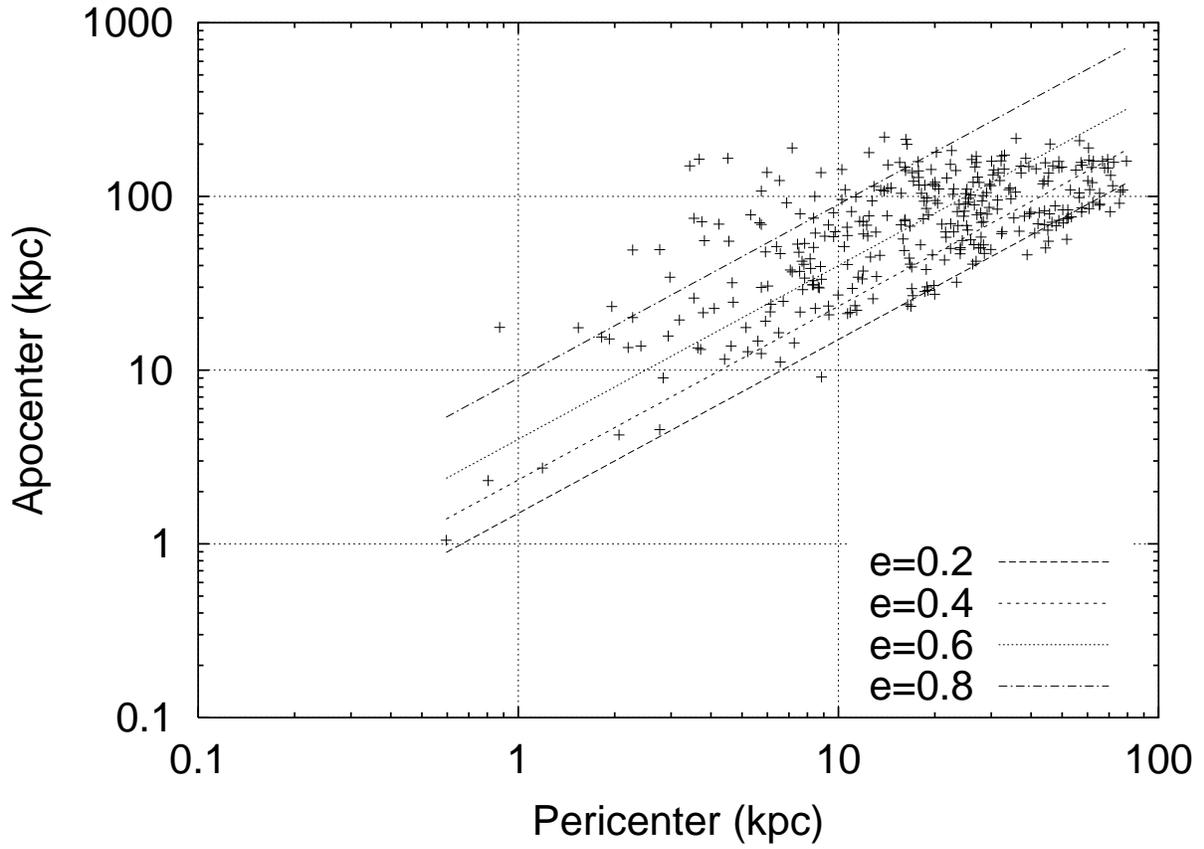}
      \caption {Distribution of the pericenter and the apocenter of the
        clumps in model~1. The cross (+) symbols indicate the
        individual clumps. The lines of constant eccentricity
        $e=0.2$, 0.4, 0.6, and 0.8 are also shown.\label{fig:dist_ecc}} 
\end{figure}

The orbital properties of the clumps could be represented by the radii
of their pericenters and apocenters. Fig.~\ref{fig:dist_ecc} shows the
distribution of the pericenters and the apocenters of the clumps. The lines of the
constant eccentricity, which is defined as $e\equiv
(r_\mathrm{a}-r_\mathrm{p}) / (r_\mathrm{a}+r_\mathrm{p})$ are
superposed in the plot. The eccentricity distribution has a
median at 0.6, which is in a good agreement with the Standard CDM
simulations that predicts high eccentricity orbits with a median
apocentric-to-pericentric distance of 6:1 (Moore et.al, 1999).

For the mass spectrum of the clumps, we examine three different
possibilities. \emph{Model 1} and \emph{model 3} have only one single population of
clumps with clump masses of $10^8 M_\odot$ and $10^9 M_\odot$,
respectively. These two models are meant to be extreme cases. 
\emph{Model 2} has a mass spectrum with a mass range of $10^8$ to $10^9
M_\odot$ according to the probability distribution
\begin{equation}
  N(M)dM \propto M^{-2} dM,
\end{equation}
following the cosmological prediction by \citet{moor1999},
\citet{klyp1999b} and \citet{ghig2000}. 
This model is considered to be the most realistic
examination. In this model the masses are assigned randomly to the
clumps. These three models were designed in order to examine the role of
the individual mass of the clumps on the disk heating.

Two additional physical effects on disk heating are studied. One
is the initial disk thickness. Thinner disks have smaller vertical
velocity dispersions, they should therefore be more sensitive to disturbances
from the clumps, and might experience a large thickening. Therefore we
ran \emph{models 4} and \emph{5} which have an initial disk scale height
of 525~pc and 1050~pc, respectively. Both models have the same mass spectrum as
 model~2.

\begin{figure}
      \plotone{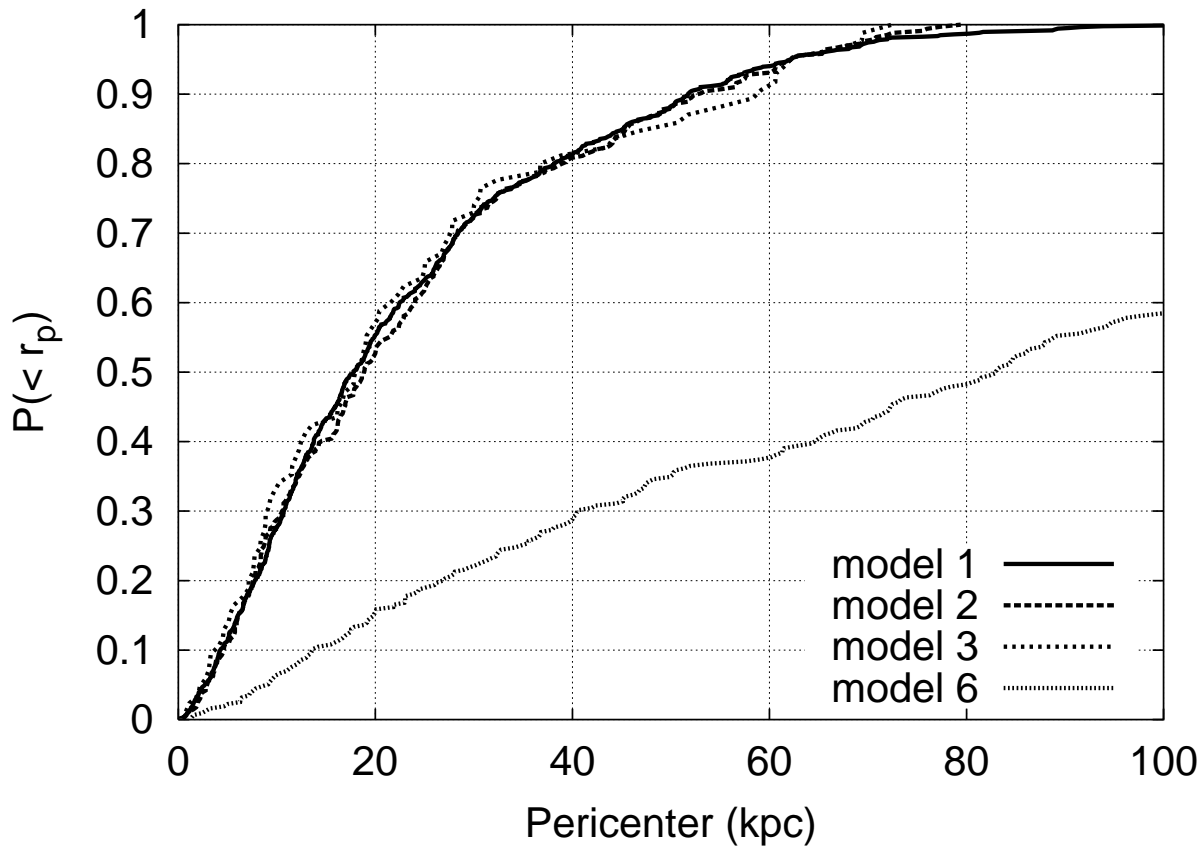}
      \caption {Cumulative distribution of the pericenters of the clumps in the
        models~1, 2, 3, and 6. The ordinate shows the fraction of the
        clumps which have a pericenter smaller than the value on the
        abscissa. In the models 1 to 3, 20 \% of the clumps have
        pericenters smaller than 10~kpc, while the corresponding
        fraction is only 5 \% in model~6. \label{fig:dist_peri}}
\end{figure}

Another influential property of the clumps on disk heating is the
position of their pericenter, where the gravitational tides become the
largest. The probability distribution of the pericentric radii of
the DM clumps is shown in Fig.~\ref{fig:dist_peri}. The abscissa shows
the pericentric radii of the clumps, and the ordinate shows the
cumulative fraction of the number of clumps which have pericenters less
than the value on the abscissa. About 18 \% -- 20 \%  of the clumps have
pericenters within the solar radius. This is similar to the clump
distribution obtained by \citet{moor1999} in the Standard CDM cosmology.

In order to examine the effect of the position distribution of the
clumps, we created \emph{model 6}, which has the same mass
spectrum as model 2, but the Milky Way has the density
profile, according to the \emph{extended halo model}. This model has
two noticeable features; (i) the inner density profile is the same as the
standard halo model so that all the kinematical properties of the disk
are equivalent, and (ii) the tidal radius is 5 times larger than that in
the standard model. We distribute the clumps in the same way as the halo
particles. Thus their spatial distribution is also 5 times
larger in radius than in the other models. In this model, the total clump
mass is the same as before. Therefore, the mass fraction of the clumps to
the whole halo is 2 \%. The probability distribution of the
pericenters is shown in Fig.~\ref{fig:dist_peri}. The number of clumps
that have pericenters within the solar radius is about 3.5 \%. This
model corresponds to the clump distribution predicted from a
$\Lambda$CDM cosmology, as demonstrated by \citet{font2001}.

The characteristic parameters of all models are summarized in Table~1.

\section{Numerical Simulation}

We calculate the interactions between the disk particles by using a
hierarchical tree algorithm \citet{hern1987} with a tolerance parameter
of $\theta_\mathrm{tol}$ = 0.7. The time integration is made with the
leap-frog method and a fixed time step of 1.75 Myr. The softening
length of the disk particles is 70 pc. We use 131,072 particles in the
disk. The forces from the clumps are calculated by direct
summation. We have followed the evolution up to 4.76 Gyr.

The calculations are made on Pentium III -- Linux workstations (800MHz).
The typical calculation time is 90 seconds per time step.

\section{Results}

In this section we show the results of our numerical simulations on the
disk heating. Three different physical effects are examined here:
the mass spectrum of the clumps, the initial disk thickness, and the
spatial distribution of the clumps.

Throughout the section, we are measuring the disk heating by the
increase of the disk thickness, $\Delta z_d$, and by the change in the
velocity dispersion, $\Delta \sigma_R$, $\Delta \sigma_z$. To calculate
these quantities, the disk is stratified in concentric annuli, each of
which containing 8192 particles. The disk thickness is evaluated simply by
the mean square deviation of the $z$-coordinate of disk particles from
the center of the disk plane, $z_d(R) \equiv \langle
{z_d}^2\rangle^{1/2}$, within the annuli.

\begin{figure}
  \epsscale{0.6}
  \plotone{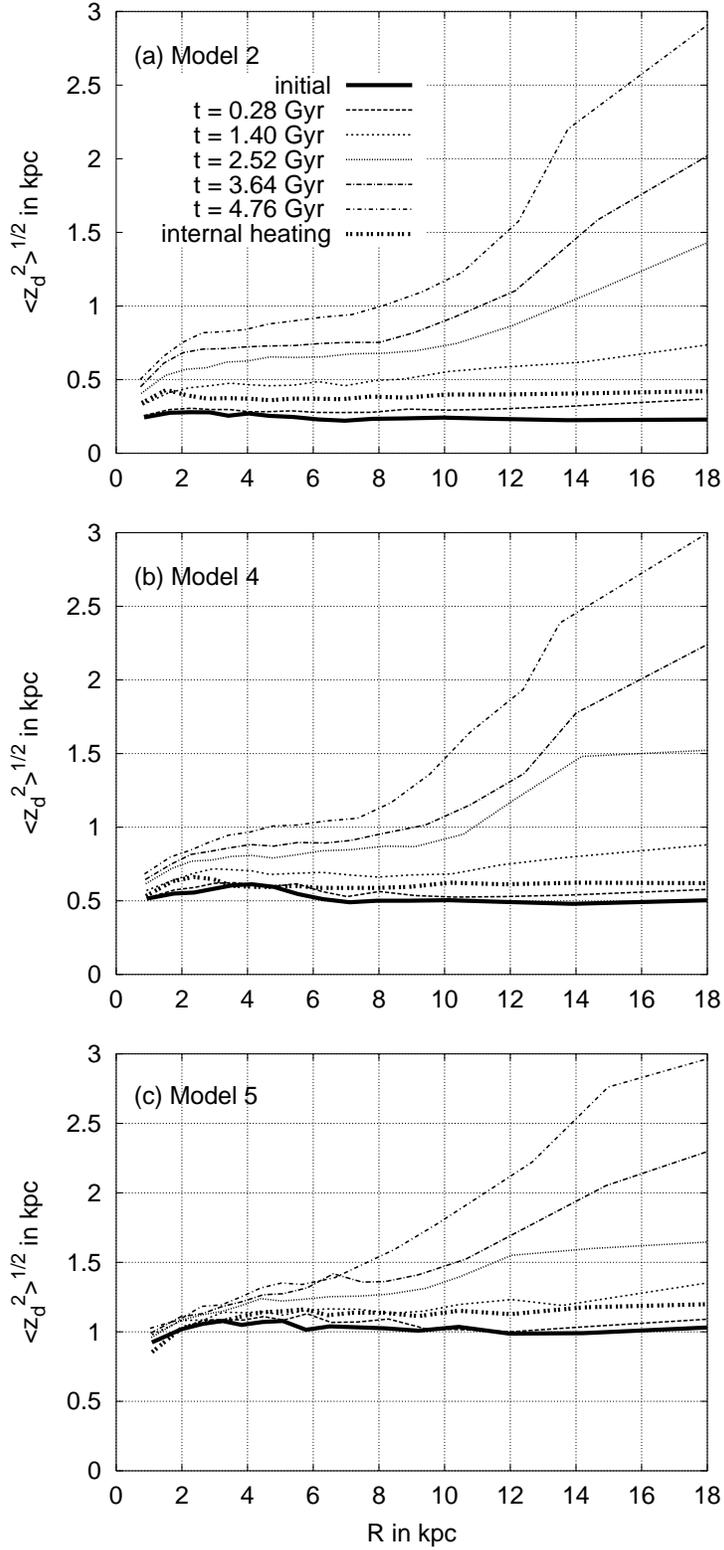}
      \caption {The growth of the disk vertical scale length which initially
      has height (a) 245 pc (b) 525 pc (c) 1050 pc, after interaction
      with DM clumps, specified in model~2, model~4 and model~5.}
  \label{fig:z245}   
\end{figure}

\begin{figure}
  \epsscale{0.6}
  \plotone{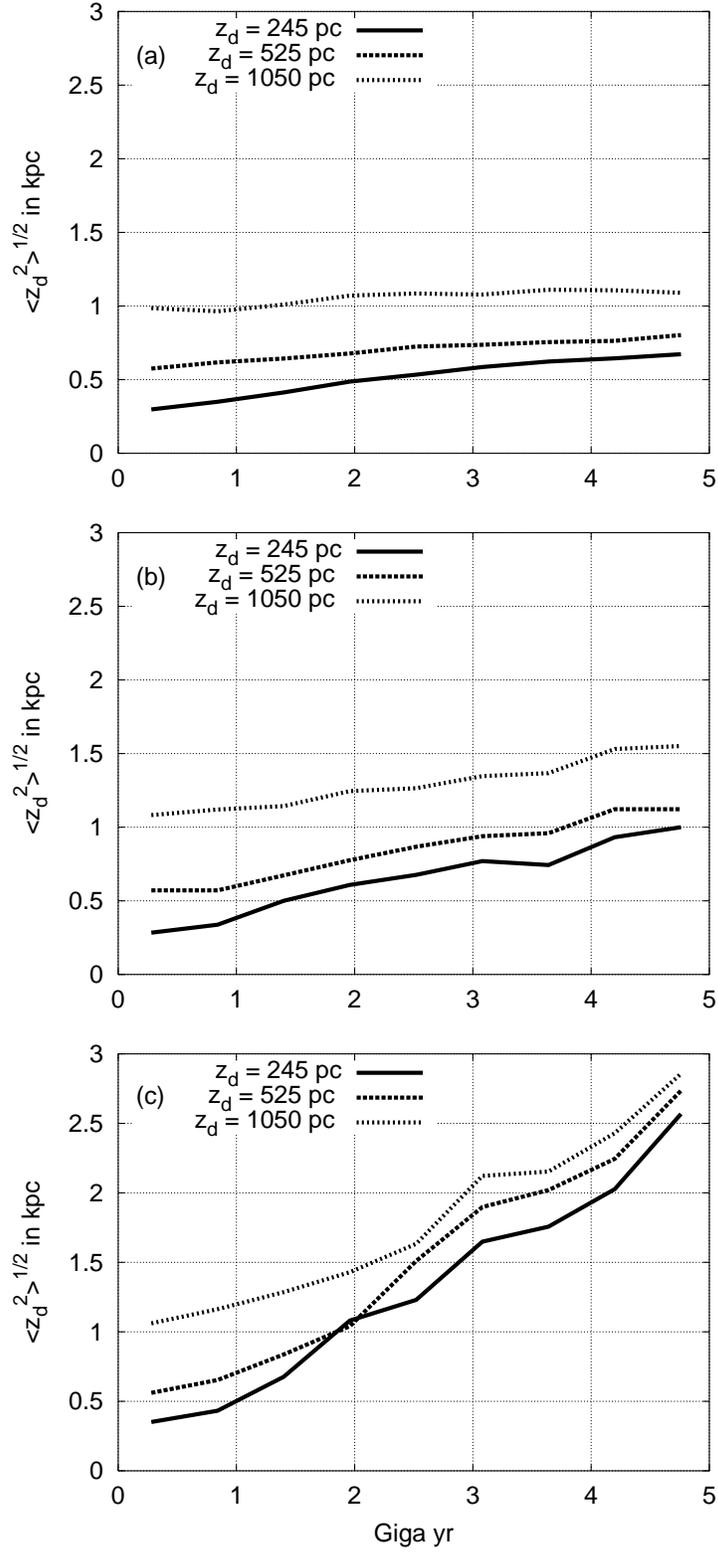}
      \caption {Heating of a disk with initial height 245 pc, 525 pc and
        1050 pc (a) at $R=2$~kpc, (b) at the solar radius, and (c) at
        $R = 16$~kpc.}
      \label{fig:zgrow_o16}   
\end{figure}

\begin{figure}
  \epsscale{0.8}
  \plotone{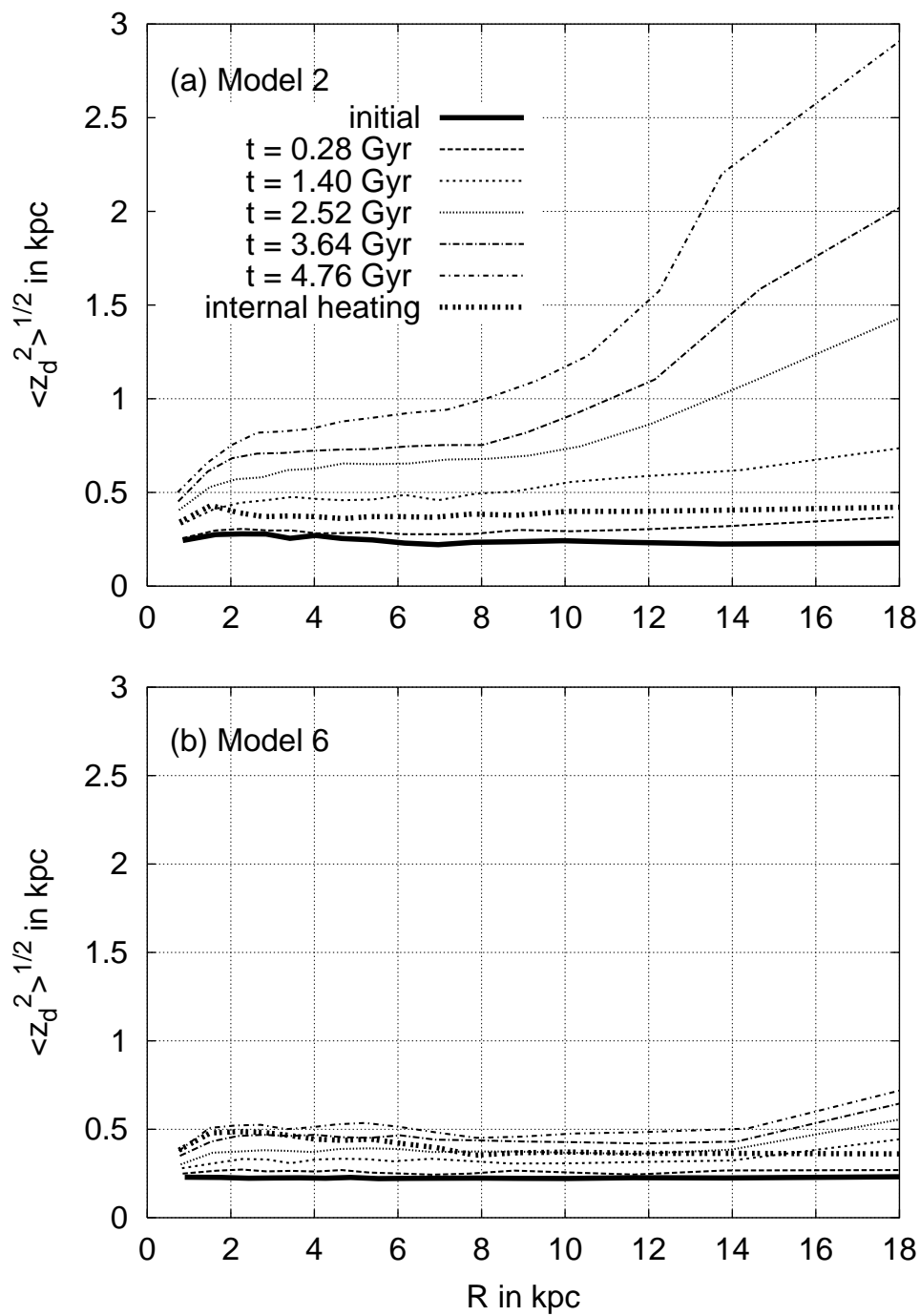}
      \caption {The growth of the disk vertical scale length produced by (a)
  model~2 where 18 \% of the clumps have pericenter radii less than the solar
  radius and (b) model~6 with only 3.5 \% of the clumps crossing the
  disk within the solar radius.}
 \label{fig:z26}   
\end{figure}

\begin{figure}
  \epsscale{1.0}
  \plotone{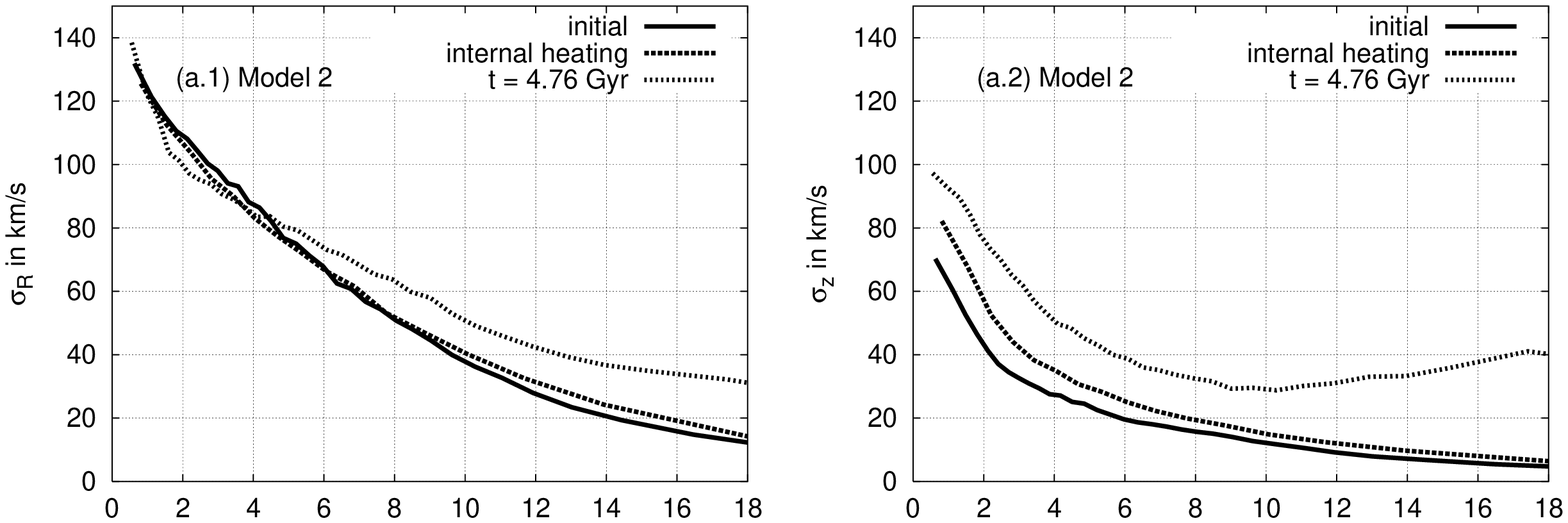}
  \plotone{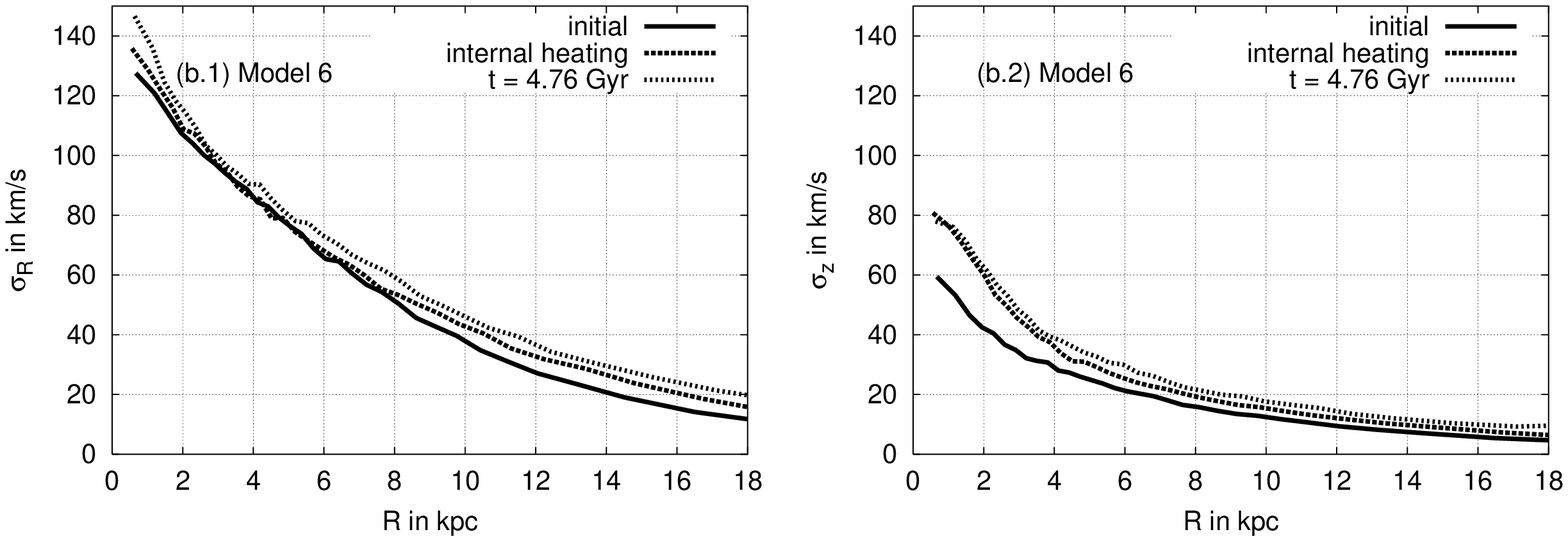}
      \caption{ Disk velocity dispersion profiles produced by models
      with a fraction of clump pericenter radii inside the solar radii
      of (a) 18 \% in model~2 and (b) 3.5 \% in model~6.}
 \label{fig:rvrz26}   
\end{figure}

Even without the clumps, the self-heating within the disk owing to 
two-body relaxation among the disk particles is always taking place in
numerical simulations with modest number of particles. In our
simulations with 131,072 particles in the disk, the increase owing to
the self-heating is typically $\Delta z_d = 150$~pc, and $\Delta
\sigma_z = 4$~km/s at the solar radius after 4.76~Gyr simulations. These
values are reasonably small, but need to be taken into account when
analyzing the results. The curves corresponding to the self-heating
after 4.76 Gyr are shown as internal heating in Fig.~\ref{fig:z1_z3} -- Fig.~\ref{fig:rvrz26} . 

\subsection{Mass Spectrum of the Clumps}

\begin{figure}
  \epsscale{0.6}
      \plotone{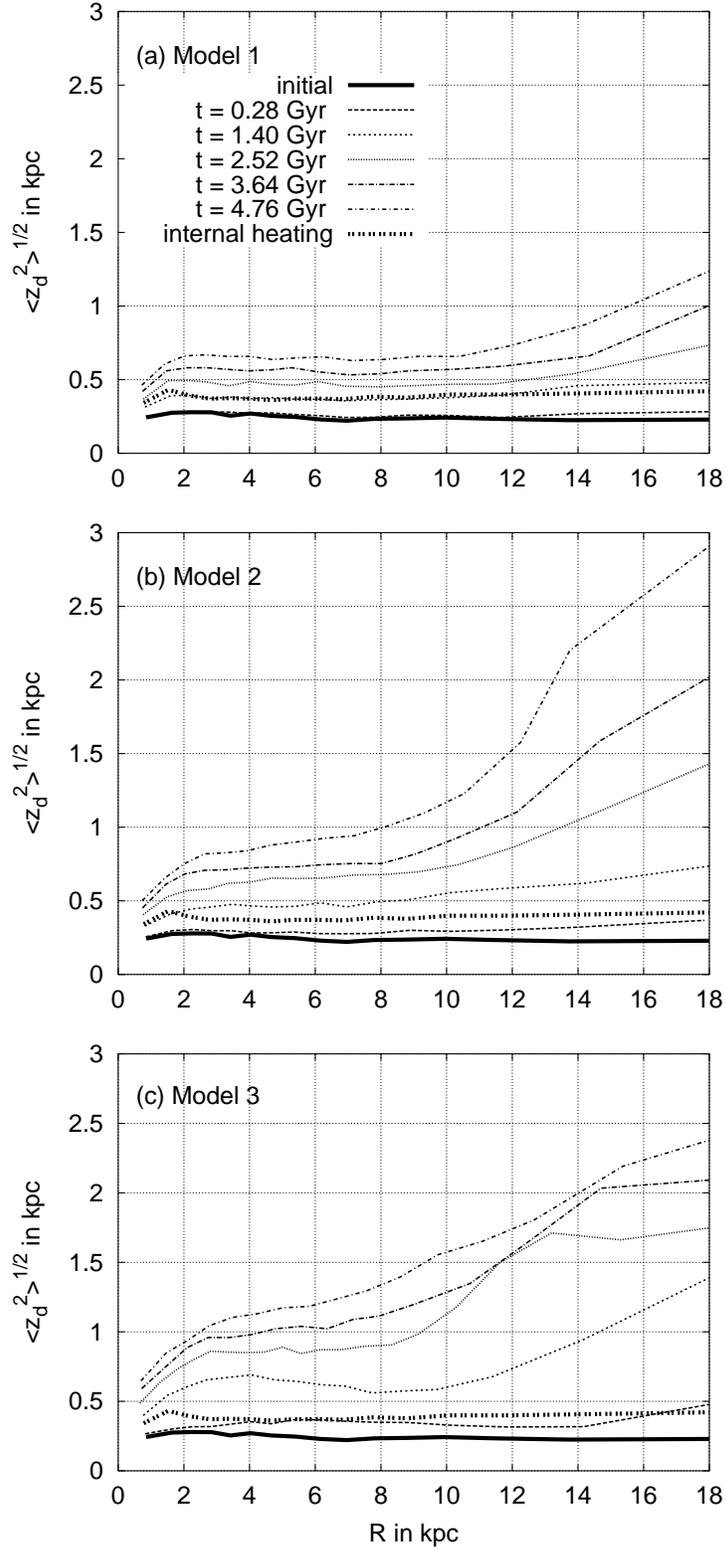}
      \caption {The growth of the disk vertical scale length after interaction
    with DM clumps with different masses given in (a) model~1 (b)
    model~2 (c) model~3.\label{fig:z1_z3}}  
\end{figure}

\begin{figure}
  \epsscale{1.0}
      \plotone{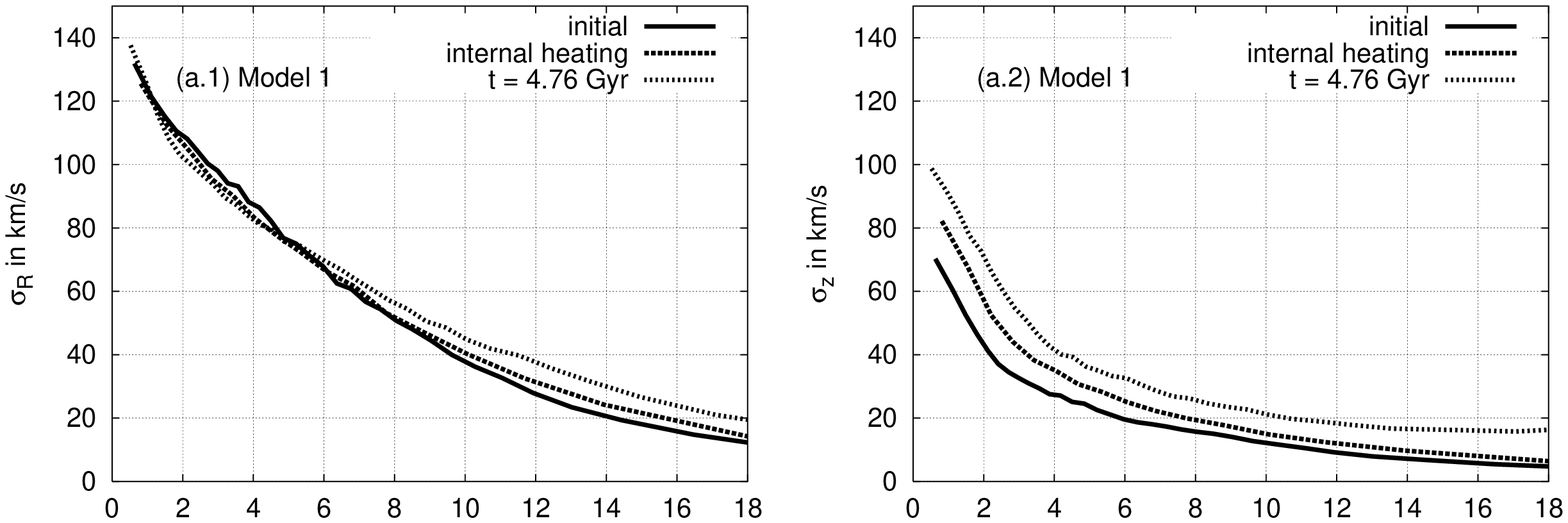}
      \plotone{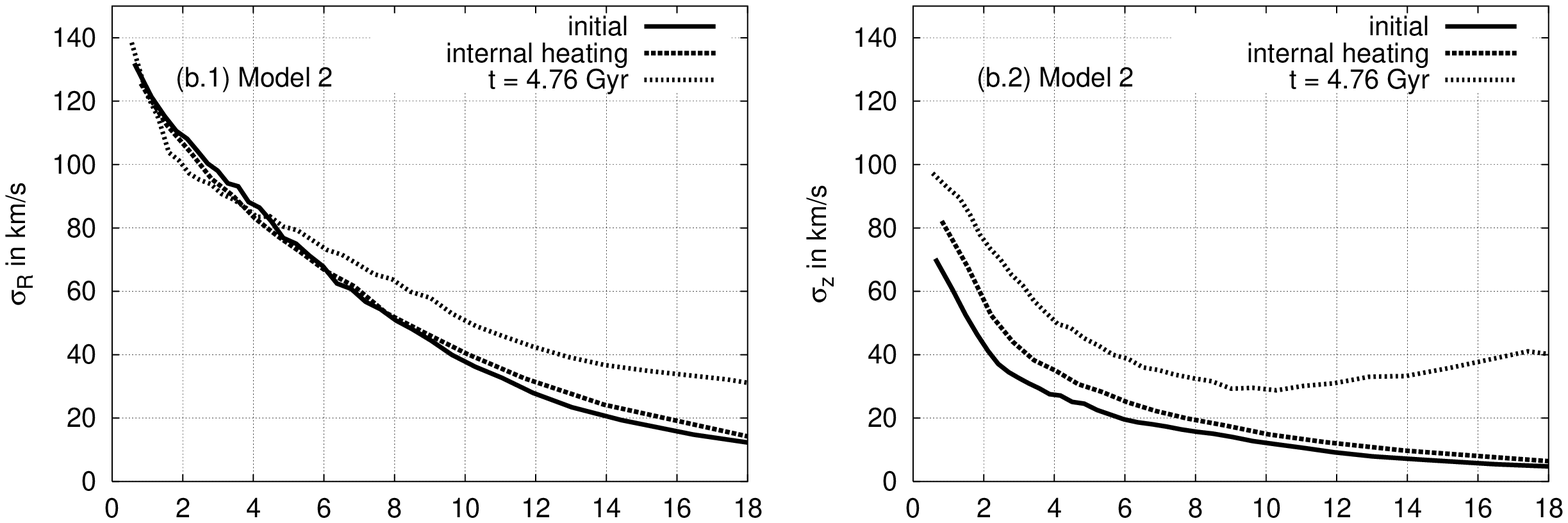}
      \plotone{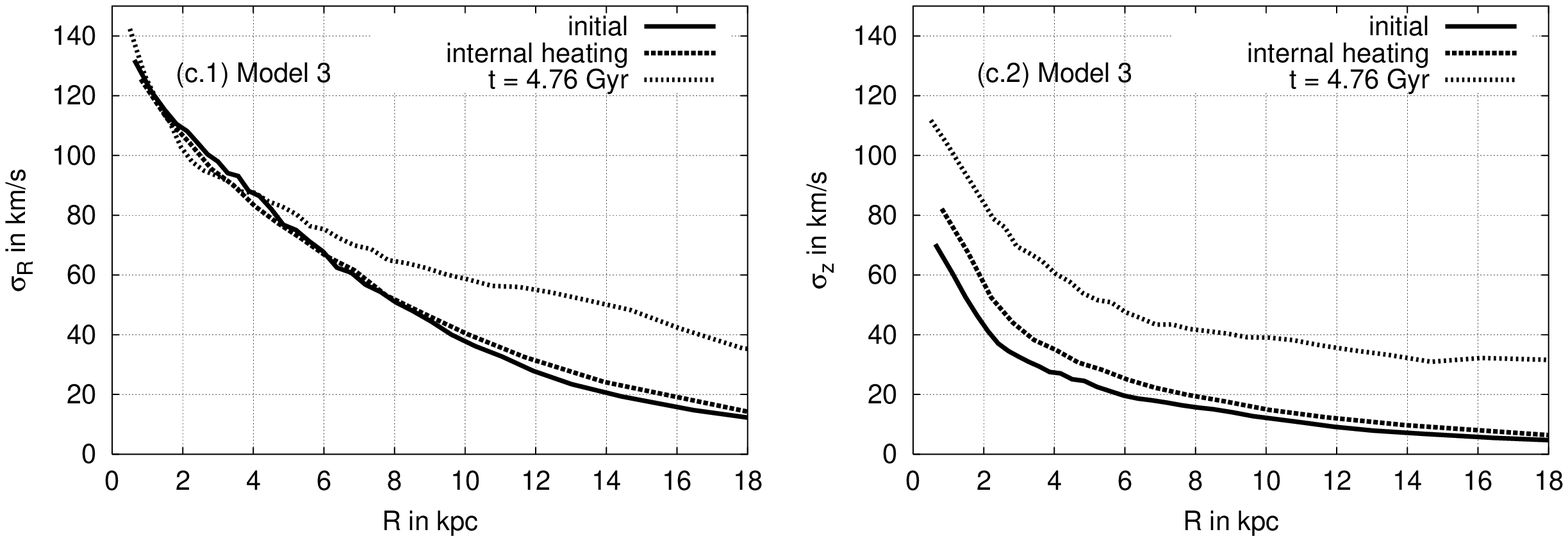}
      \caption{Disk velocity dispersion in vertical and radial direction
  after 4.76 Gyr of interaction with different mass distributions of DM
  clumps given in (a) model~1 (b) model~2 (c) model~3.}
 \label{fig:rvrz1_3}   
\end{figure}

The results of the disk heating obtained from the simulations of the
model 1, 2, and 3 are shown in Fig.~\ref{fig:z1_z3} and
Fig.~\ref{fig:rvrz1_3}. The comparison between the results makes the
effect of the mass spectrum of the DM clumps clearer.

The growth of the disk thickness in model 1 is $\Delta z_d \sim$ 615~pc,
630~pc, and 900~pc at $R=2$, 8, and 16~kpc, respectively, while those
values in model~3 are 860~pc, 1298~pc, and 2197~pc. It is clear that the
interaction of a galactic disk with a few, but massive clumps (model~3)
is more effective to heat the disk than the interaction with many but
less massive clumps. This result can be explained by the estimate in an
impulsive approximation, which predicts that the increase in the energy
of the disk is $\Delta E \propto N~M^2$. If there is a mass spectrum, $d
N(M) \propto M^{-2}d M = M^{-1} d(\log M)$, $\Delta E$ is dominated by
massive clumps. The disk thickening obtained from model~2
(Fig.~\ref{fig:z1_z3}\textbf{b}) is, in fact, much closer to that in model
3. This result could be also understood by the impulsive approximation,
as the energy input to the disk from the logarithmic interval in the
mass spectrum is dominated by the more massive clumps; $d\Delta E
\propto M d(\log M)$. Therefore the higher limit of the mass spectrum
has an essential contribution on the disk heating.

A difference feature between models 2 and 3 is their dependence of the
disk heating on the disk radius. For all models the outer part of the
disk is heated up more than the inner part, but this gradient is the
largest in the model 2. The increase of the disk thickness for $R<13$~kpc
is smaller in model 2 than in model 3, while it is opposite in
the outer disk. In model 2, which imitates the results of the
Standard CDM model, the disk thickness becomes 1~kpc at the solar radius
after 4.76~Gyr interactions.

An alternative way to estimate the disk heating is the increase in the
velocity dispersion. Fig.~\ref{fig:rvrz1_3} shows the same results as
Fig.~\ref{fig:z1_z3}. The existence of a few massive clumps has a large
influence on the disk heating. This figure shows the increase in the
radial (the left panels) and the vertical (the right panels) velocity
dispersions. After 4.76 Gyr, the interaction of the disk with clumps in model 2
produces $\sigma_R = 60$ km/s and $\sigma_z= 35$ km/s. The properties of
disk heating in model 2 seem to agree with those of the thick disk
component at the solar radius which has  $\sigma_R = 63$ km/s and
$\sigma_z= 38$ km/s and a vertical scale length of 1~kpc \citep{free1996}. 

It is noticeable that the initial velocity dispersion is not
isotropic or $\sigma_R$ is larger compared to $\sigma_z$. After 4.76
Gyr, in the inner part of the disk ($R<5$~kpc), there is no increase in
the radial velocity dispersion while the vertical velocity dispersion
increases a lot. This anisotropy in the heating is seen also in the
outer part of the disk in model 2. A possible explanation is that the
heating owing to the clumps is aiming at a more isotropic velocity dispersion.

\subsection{Initial Disk Thickness}

\begin{figure}
  \epsscale{1.0}
  \plotone{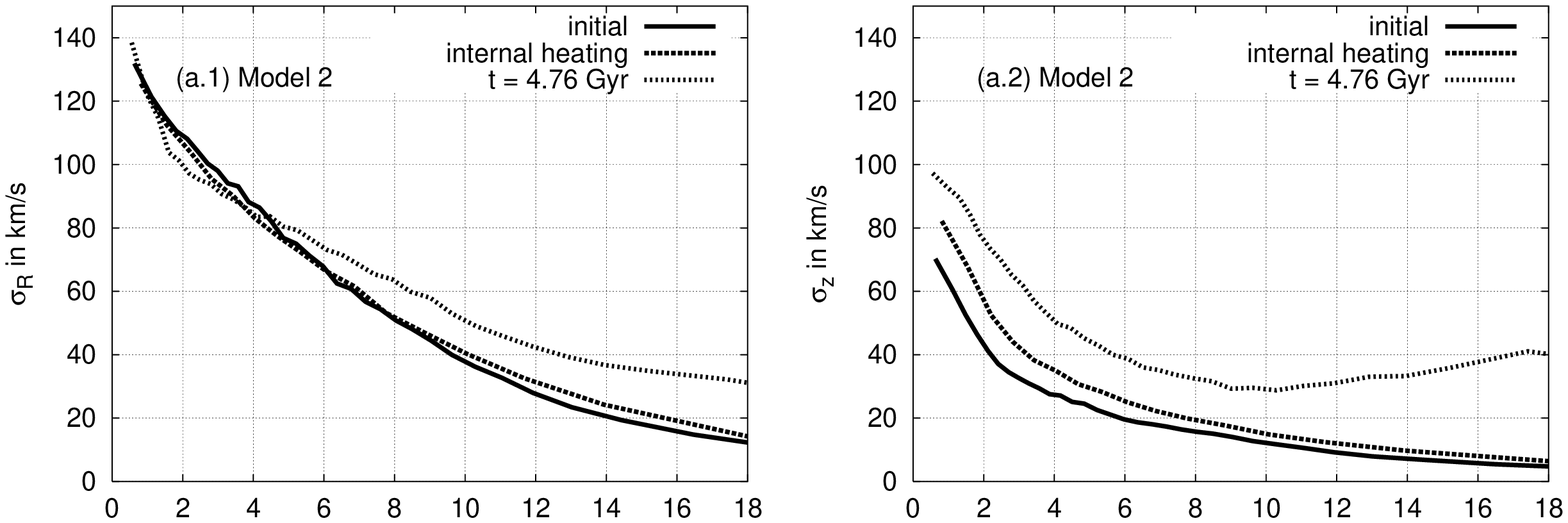}
  \plotone{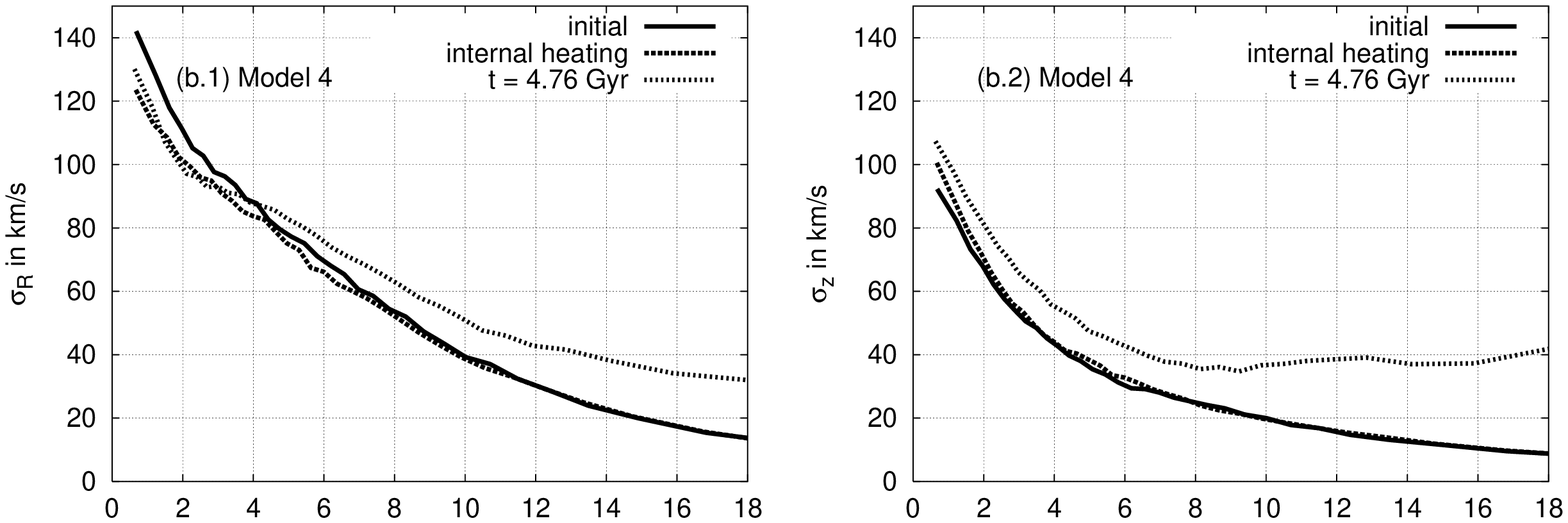}
  \plotone{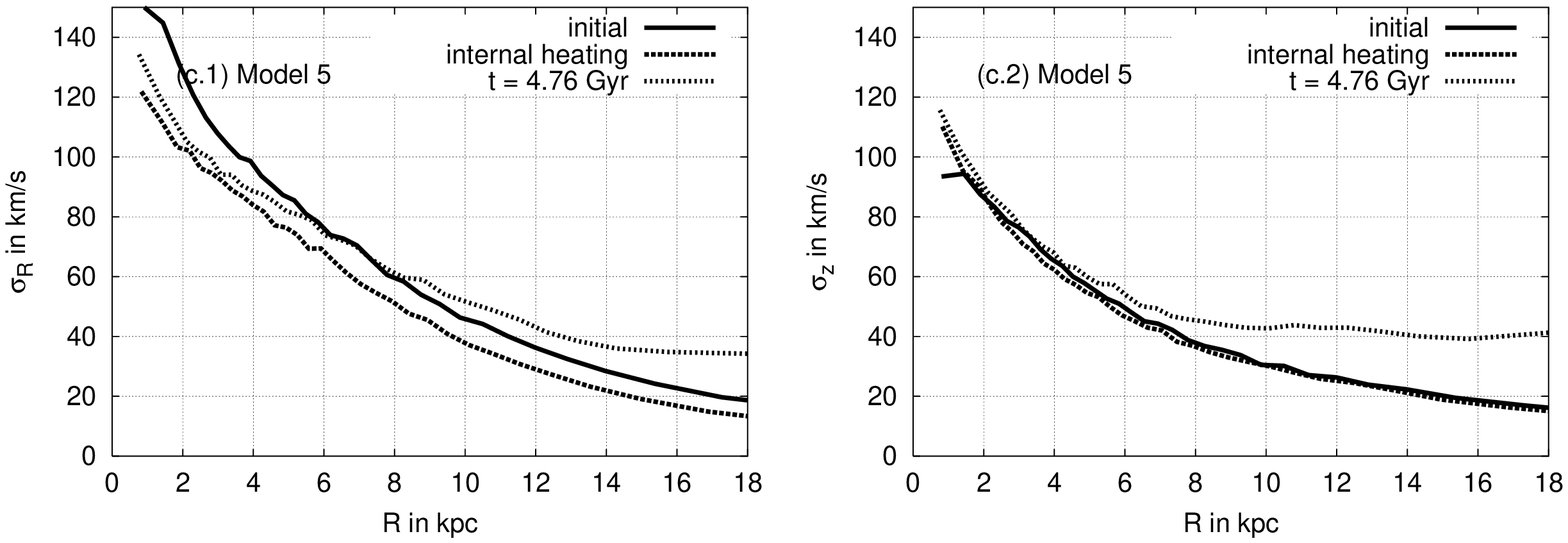}
      \caption{Velocity dispersion of a disk which initially has a
      height of (a) 245 pc (b) 525 pc (c) 1050 pc, after 4.76 Gyr of
      interaction with DM clumps, specified in model~2, model~4 and
      model~5.}
     \label{fig:rvrz245}   
\end{figure}

Our model 2 has an initial disk thickness comparable to the observed
value in the Milky Way, which is about 250 pc. In many simulations in
other works, however, the typical initial disk thickness was chosen much
larger. For example, \citet{font2001} employ the initial disk with a
thickness of 700 pc. The susceptibility of the disks to the heating owing
to the interaction with the DM clumps should depend on the initial disk
thickness. Thicker disks are possibly less sensitive and suffer less
heating. In order to clarify the difference in the disk heating
according to the initial disk thickness, we carried out three investigations
with the models 2, 4, and 5. These models have almost the same bulge, halo, and
clump systems, but the initial disk thicknesses are 245~pc, 525~pc, and
1050~pc, respectively. The results of the disk heating are shown in
Fig.~\ref{fig:z245}, and Fig.~\ref{fig:rvrz245}.

After 4.76 Gyr, the disk in model 2 achieves a thickness of 1~kpc at
the solar radius, thus the increase rate is about 400\%. In models 4
and 6, the final disk thicknesses at the solar radius are 1.1~kpc and
1.5~kpc, respectively. The increase rates in the thickness are thus
210\% and 70\%. This tendency is more pronounced in the inner regions. In
the central region $R<3$~kpc, the disk in model 5 has experienced
nearly no heating. On the other hand, in the outer regions $R>14$~kpc,
the heating owing to the interaction with the clumps is so large that
the difference in the initial disk thickness is negligible. This feature
is more clearly seen in Fig.~\ref{fig:zgrow_o16}. The panels in the
figure show the time variations of the disk thickness (a) at $R=2$~kpc,
(b) at the solar radii, and (c) at $R=16$~kpc, respectively. It is
natural that the outer part of the disk experiences more heating,
because it is closer to the DM clumps. A common characteristic to all
radii is that the difference in the initial disk thickness decreases in
time. The curves in the panel (a) indicate that the heating will be
saturated at the level of $z_d\sim 1.2$~kpc. In the outer parts the
heating is still going on to make the disk even thicker. The growth of
the velocity dispersions in the disks, which is shown in
Fig.~\ref{fig:rvrz245}, also gives the same information namely that
thicker disks suffer less heating.

This flaring effect, that is the increase in the disk thickness
especially in the outer part, is seen in VLA observations of HI
disk of galaxies NGC 100 and UGC 9242 \citep{bosm1991}. The interaction
with the hypothetical DM clumps would be an explanation of this feature.

\subsection{Spatial Distribution of the Clumps}

In our models the clump's spatial distribution is the same as the halo
distribution in order to ensure the equilibrium. This clump
distribution, however, creates more clumps in the inner part of the halo
than that obtained by the cosmological simulations. For example,
Font et al (2001) employ a clump distribution in which only 3 \% of
the clumps have pericenters within the solar radius. Our model 6 is
designed to remove the clumps from the inner region. This is accomplished by
introducing a more extended equilibrium halo model. We distribute the
same clumps as in the model 2 in the extended halo. The spatial
extent of the clump distribution in this model becomes 5 times larger,
and only 3.5 \% of the clumps have pericenter within the solar
radius (see Fig.~\ref{fig:dist_peri}).

The increase in disk thickness and velocity dispersion is shown
in Fig.~\ref{fig:z26}~and~Fig.~\ref{fig:rvrz26}. In the model 6, the
disk heating owing to the interaction with the clumps is comparable with
the disk internal heating, up to the disk edge. In this case, we would need
more precise simulations with larger number of particles, but it is
already clear that even with the presence of the many clumps surviving
in the halo, tidal disk heating is negligible
and the disk remains as thin as the real Milky Way disk for about 5
Gyr. This result is in agreement with conclusion of \citet{font2001}.

\section{Discussion and Conclusions}

We have made a series of numerical simulations, examining heating of 
galactic disks that are embedded in a halo with many small DM clumps, as
suggested by cosmological simulations. We have built up the initial galaxy
model as precisely in equilibrium as possible, so that we could simulate
a stable disk that is as thin as the Milky Way.

We have shown that with the mass spectrum $N(M)dM \propto M^{-2}dM$,
that is consistent with cosmological simulations, the massive end of
the mass spectrum determines the amount of disk heating. The number
of the clumps at the massive end could be only a few, so that the massive
limit of the spectrum might fluctuate among galactic halos. This fact
suggests a variety of disk thickness because of the random
fluctuation in the number of the most massive clumps.

As a second result, the simulations demonstrate that thicker disks
suffer less heating. It is possible that a disk with 700~pc thickness
shows no significant heating whereas another disk with 200~pc thickness suffers
significant heating in the same circumstances. This also means that
there is a critical thickness at which the heating owing to the
interaction with the clumps is saturated. 

Finally, we have considered the relation to cosmology. Our model 2 mimics
the clump distribution found in the Standard CDM cosmology
\citep{moor1999,klyp1999b}. The disk heating by the clumps has already
been discussed by \citet{moor1999}. They argued by a simple impulsive
approximation that disk heating would be too high to explain the
age-temperature relation for disk stars \citep{wiel1974}. Our numerical
simulations are supplementary to their analysis, and we confirm
that the heating is significantly high. The thickness of 1~kpc at the
solar radii is comparable to the present thick disk, but this heating
takes place within 5~Gyr and even a young thin disk would have no chance to
survive. On the other hand, our model 6 corresponds to the clump
distribution of the $\Lambda$CDM cosmology, similar to the one
\citet{font2001} have studied, but more closely representing the Milky
Way. As found by \citet{font2001}, this clump
distribution causes no significant heating of the disk. These results
suggest that the $\Lambda$CDM cosmology is preferable with respect to
the Standard CDM cosmology in explaining the thickness of the Milky Way.

\acknowledgments

We thank Burkhard Fuchs and Andreas Just for their useful comments and
suggestions. EA is grateful to Astronomisches Rechen-Institut (ARI),
Heidelberg, in particular Rainer Spurzem, for allowing her to conduct
simulations on ARI's Unix cluster. TT is indebted to Alexander von Humboldt
foundation for research grant.

\clearpage

\clearpage

\begin{deluxetable}{crrr}
\tablecaption{DM clumps and disk models.\label{tab:models}}
\tablewidth{0pt}
\tablehead{
\colhead{Model} & \colhead{Mass of each clump} & 
\colhead{Number of clumps} & \colhead{Disk height}
}
\startdata
Model 1 & $10^8$ $M_\odot$ & 859 & 245 pc \\
Model 2 & $10^8 - 10^9$ $M_\odot$ & 335 & 245 pc \\
Model 3 & $10^9$ $M_\odot$ & 86 & 245 pc \\
Model 4 & $10^8 - 10^9 $ $M_\odot$ & 335 & 525 pc \\
Model 5 & $10^8 - 10^9 $ $M_\odot$ & 335 & 1050 pc \\
Model 6 & $10^8 - 10^9 $ $M_\odot$ & 315 & 245 pc \\
\enddata
\end{deluxetable}

\end{document}